# Analysis of Skin Effect in High Frequency Isolation Transformers

Mohamad Saleh Sanjarinia, Sepehr Saadatmand, Pourya Shamsi, and Mehdi Ferdowsi
Department of Electrical and Computer Engineering
Missouri University of Science and Technology
Rolla, Missouri, USA
Corresponding Author: Mohamad Saleh Sanjarinia (mswvq@mst.edu)

*Abstract*— **In this paper, a high frequency transformer with different conductors and winding arrangements, at presence of eddy currents and skin effect, is studied. By using different winding structures, and conductor types, such as circular, square shaped, and foil wires, the skin effect in the windings is studied and current density within the conductors at a high frequency of 20 MHz and a lower frequency of 20 kHz are investigated using finite element method (FEM) simulation. Moreover, magnetic field distribution in the transformers at 20 MHz is obtained and displayed. Also, magnetizing inductance, leakage inductance and AC winding resistance for all of the transformer types are found and compared, and frequency response for the transformers are obtained and shown. Lastly, based on the results, the skin effect increases the AC winding resistance and decreases the leakage inductance as the frequency increases. Furthermore, different winding arrangements, conductors, and transformer types show a wide range of parasitic and loss behavior, which enable the designers to compromise between various parameters in different applications, especially new fast switches such as SiC and GaN.**

## I. INTRODUCTION

SWITCHING frequency is an important factor in determining the size of the passive components, in a power electronics converter. Objectives such as miniaturization, cost reduction, and ease of manufacturing have resulted in the converters with higher switching frequencies [1]. In spite of some advantages such as better transient performance and improved integration of passive components at higher frequencies, achieving a reliable and an efficient high frequency converter, is still challenging [1,2]. Although, a higher switching frequency can reduce the overall volume of magnetic elements, it will add to the core/copper losses and will generate further electromagnetic interference (EMI) especially at high frequency (HF) (3 MHz < $F_{sw}$ < 30 MHz) and very high frequency (VHF) (30 MHz < $F_{sw}$ < 300 MHz) applications [2]. Leakage inductance and AC resistance of the windings can significantly impact the performance and efficiency of the transformer.

In order to reduce switching losses in the HF converters, zero voltage switching (ZVS) and zero current switching (ZCS) are commonly used. In such scenarios, if the leakage inductance and parasitic capacitance of the inductors are not considered and accurately estimated, additional modes of oscillation can appear in the circuit, and can lead to lower efficiencies as well as creating problems in control systems, especially in new control techniques for power electronics converters [3-5]. However, if these parameters are known, they can be incorporated in the ZVS/ZCS resonant tanks to reduce the overall size of the converter [6]. Leakage inductance is directly related to the winding topology and transformer structure. Therefore, to see the effectiveness and impact of the transformer structure on the values of leakage inductance, analyzing transformers with different winding arrangements, considering the skin effect, can be beneficial.

To obtain an accurate estimate of the leakage inductance and AC resistance in HF/VHF applications, it is necessary to consider the skin effect. Because of the magnetic field generated by the passing current, the current is pushed towards the surface of the conductor. This phenomenon is called skin effect and leads to a lower utilization of the inner surface of the conductor; and consequently, a smaller effective current carrying cross section area. As the frequency goes higher, the skin effect intensifies. Therefore, the cross-sectional area decreases. As the inductance value is also related to the cross sectional area, the leakage inductance decreases. Moreover, the resistance is inversely proportional to the effective area of the cross section. Thus, by increasing the frequency, it is expected to see higher resistance in the windings. Hence, inclusion of the skin effect in the analysis of HF inductors is crucial.

In this paper, a transformer with different winding arrangements and different conductor types, were designed and considered at 20 MHz. By considering the skin effect, the magnetic fields are depicted by using FEM simulations. Then, leakage inductance and winding AC resistance are obtained and compared with each other. The results show a range of leakage inductance and AC resistance, which can be used in VHF applications. Also, comparing different structures can help to understand, which topology and structure is better for which application.

## II. TRANSFORMER MODEL AND STRUCTURE

In this section, the transformer topology, which used in this study is displayed. The transformer dimensions are selected, based on the procedure discussed in [1,7]. As the frequency increases, the passive component size decreases, but core loss may limit further decrease in the transformer size. Equation (1) shows the relation between cross sectional area and frequency.

$$A = \frac{E}{KNfB} \tag{1}$$

In equation (1), $A$ is the cross sectional are of the core, $E$ is the RMS voltage produced by the flux change over time based on the Faradays law, $K$ is the waveform constant, $N$ is number of turns, $f$ is the frequency, and $B$ is amplitude of flux peak.

By using Steinmetz equation loss density of the core can be found.

$$P = K \cdot f^{\alpha} \cdot B^{\beta} \tag{2}$$

In this equation $P$ is the average of the power loss density, $B$ is the amplitude of the flux peak, $f$ is the frequency (in kilohertz), and $K$, $\alpha$ and $\beta$ are the constants of magnetic core.

Also, winding loss can be found by using equation (3). As the very high frequency is the goal, involving skin depth as a parameter which show the depth of penetrating current into the conductor, is very influential.

$$P = \frac{1}{2} I_{ac}^2 R_{ac} = \frac{1}{2} I_{ac}^2 \frac{\rho \iota_\omega}{\omega_\omega \delta} = \frac{1}{2} I_{ac}^2 \frac{\rho}{\omega_\omega} \sqrt{\pi \rho \mu_0 f} \tag{3}$$

In equation (3), $I_{ac}$ is the sinusoidal ac current amplitude, $\rho$ is the conductor resistivity, $\iota_\omega$ is the length of the winding, $\omega_\omega$ is the effective width of the conductor, and $\delta$ is the skin depth. It is clear that, the skin depth is inversely proportional to the frequency value.

The VHF transformer studied here, is an EE core with ferrite-NiZn 67-Material core type, which B-H curve of the core used for simulations are depicted in Fig. 1. The ferrite is a suitable core for high frequency applications. Low core losses, low relatively cost, and good temperature stability are some of the features of the ferrite cores. 67-Material is a low loss magnetic component that is suitable for very high frequency applications (up to 50 MHZ). But most of its applications are between 2-20 MHz. Over 20 MHz lack of material performance data is a challenging issue for designing magnetic material components. For the core loss, equation (2) for the 67-material at 20 MHz can be displayed by the equation (4) (at duty cycle of 0.5). The core loss at three peak flux density, at this frequency is presented in TABLE I.

$$P = 8.2060 B^{2.0959} \tag{4}$$

TABLE I
Constant values of the core for the Steinmetz equation [7].

| Flux density peak (mT) | Core loss (mW/cm³) |
|---|---|
| 2.722 | 66.652 |
| 3.869 | 141.144 |
| 5.445 | 284.903 |

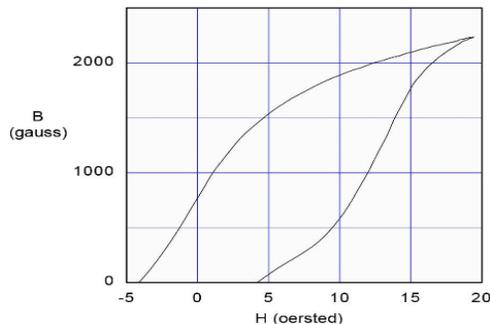

Fig. 1. Ferrite-NiZn 67-material B-H curve loop used in the simulations [8].

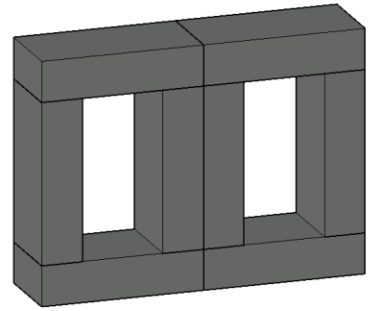

Fig. 2. The 3D EE transformer core

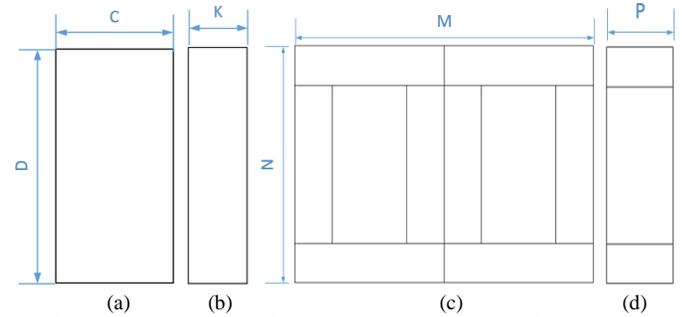

Fig. 3. Side and front views of the EE core. a) Front view of the block set. b) Side view of the block set. c) Front view of the EE core. d) Side view of the EE core. Side views show the depth of the cores.

TABLE II
Dimensions of the Fig. 3 for the EE core.

| C | 3 (mm) |
|---|---|
| D | 6 (mm) |
| K | 1.5 (mm) |
| M | 12 (mm) |
| N | 6 (mm) |
| P | 3 (mm) |

The EE core is composed of eight block sets, which is clear in Fig. 2. TABLE II and Fig. 3 show dimensions of each block set and the EE core of the transformer.

The transformer in this paper is working at 5V and 10W. Primary and secondary have 5 turns on each. The transformer ration is 1:1.

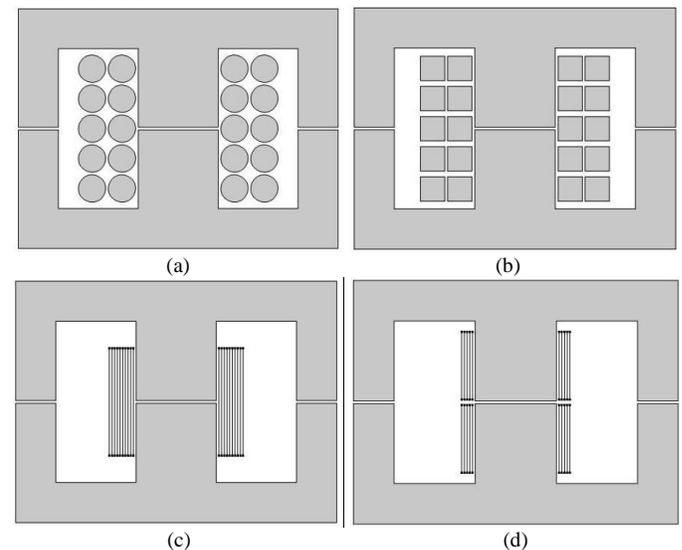

Fig. 4. Different winding arrangements of the transformers (all equivalent to the AWG18). a) Windings with the circular conductors b) Windings with the square shape conductors the with same cross sectional area to the AWG18. c) Type (I) of the windings with foil conductors. d) Type (II) of the windings with foil conductors. Each of the airgaps in the cores are 0.1mm.

To investigate the leakage inductance and AC resistance, with respect to the skin effect, in very high frequency, four types of winding arrangements are considered. The winding arrangements are displayed in Fig 4. For this purpose, circular and squared shape wires equal to the AWG18 as well as a foil with the equivalent cross sections have been used. For the foil winding, both overlaid and non-overlaid patterns are investigated.

### III. Field Distributions, Skin Effect, and Leakage Inductance Calculations

In this section, results of the FEM simulations for the transformer arrangements are displayed. To show the skin effect, current density (J) distribution during the full load working of the transformers for circular and square shapes are obtained. To calculate the leakage inductance, by short circuiting the secondary and injecting nominal current to the primary the magnetic field strength (H) are depicted. For the EE cores, the leakage inductance can be found by calculating magnetic energy between two window areas, while the secondary is short circuited. Equation (4) shows the relation between magnetic energy and leakage inductance:

$$L_{leakage} = \frac{\oiiint H.B dv}{I_{primary}^2} \quad (5)$$

In equation (5), $v$ is the volume of the space that energy is calculated on, $\mu$ is the permeability of medium, $L_{leakage}$ is the leakage inductance observed from primary side, and $I_{primary}$ is the current of the primary winding.

#### A. Field Distribution and Skin Effect

Fig. 5 shows the current density distribution in the circular and the square wires during the full load working of the transformer. As it is clear, in center parts of the wires, the current density is very low, but at the parts near the surface the current density is much higher.

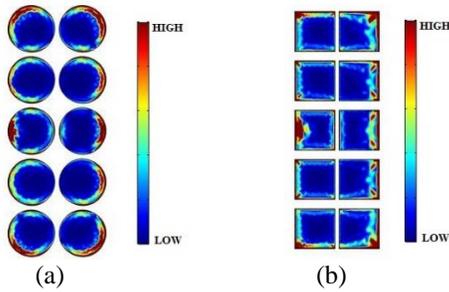
(a)      (b)

Fig. 5. Current density distribution (J) in the windings of the transformer for type (a) and (b) of the Fig. 4 at 20 MHz. a) AWG18 circular shape conductors of the primary and the secondary, at the right window. b) Windings with the square shape conductors of the primary and the secondary with the same cross sectional area to the AWG18 at the right window.

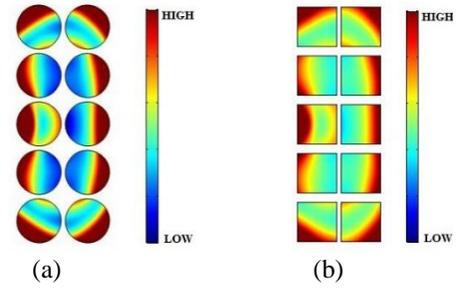
(a)      (b)

Fig. 6. Current density distribution (J) in the windings of the transformer for type (a) and (b) of the Fig. 4 at 20 KHz. a) AWG18 circular shape conductors of the primary and the secondary at the right window. b) Windings with the square shape conductors of the primary and the secondary with the same cross sectional area to AWG18 at the right window.

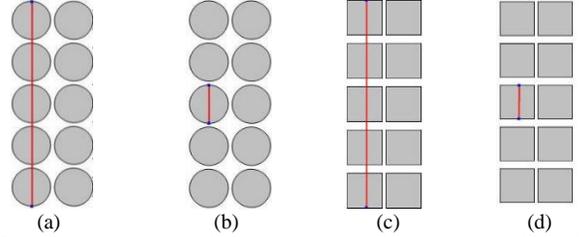
(a)      (b)      (c)      (d)

Fig. 7. Defined cutline across the conductors on the primary at the right window.

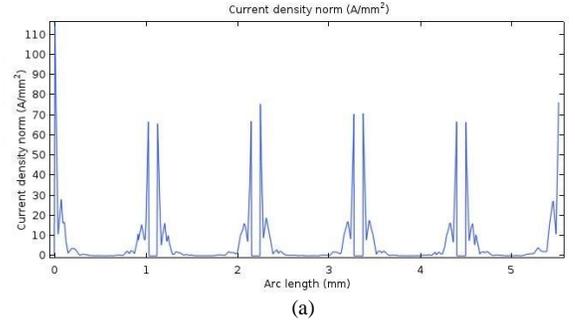
(a)

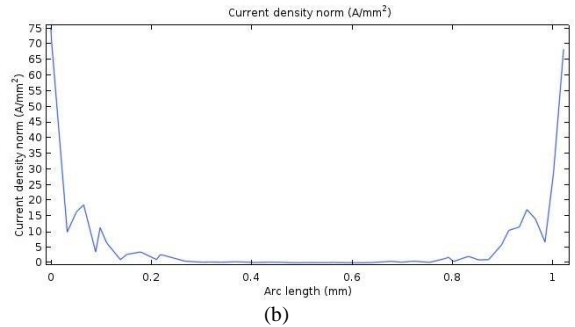
(b)

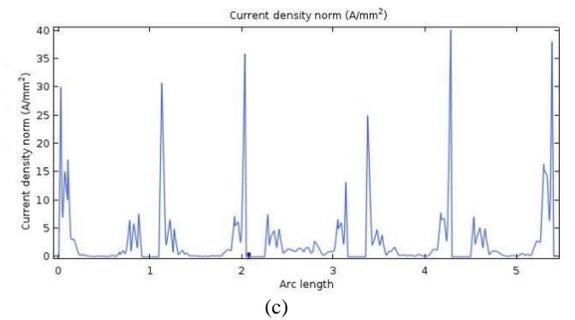
(c)

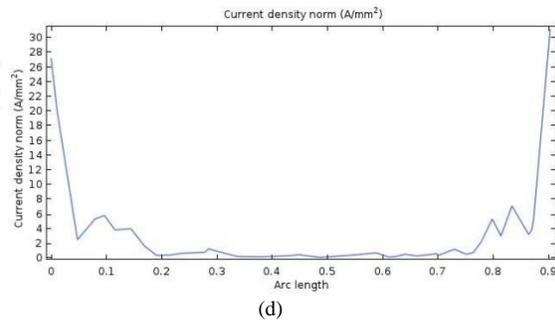

(d)

Fig. 8. Current density norm (A/mm$^2$) across the defined cutline in Fig. 7 from bottom to above at 20 MHz. a) Current density on the cutline of Fig.7 (a) across the primary conductors at the right window. b) Current density on the cutline of Fig.7 (b) across the primary conductors at the right window. c) Current density on the cutline of Fig.7 (c) across the primary conductors at the right window. d) Current density on the cutline of Fig.7 (d) across the primary conductors at the right window.

To show the influence of the skin effect due to higher frequencies, current density of the transformer for both mentioned windings are depicted, at the frequency of 20 kHz in Fig. 6. In Fig. 6 it is clear that, the current penetrated much deeper inside wires than Fig. 5 which shows J at 20 MHz. On the other word, the skin depth difference, which is clear equation (3) at higher frequencies is much lower than lower frequency. It is expected that the skin effect decreases the leakage inductance and increases the AC resistance of the windings, because of lowering the cross sectional area in the conductors. Moreover, to show how the current density values change in the conductors, at very high frequency, and see the numeric variations of the skin effect at 20 MHz, by defining the cutline that are shown in Fig. 7 the current density across the conductors are displayed at the graphs of Fig. 8. It is clear in Fig. 8 (a) and (c) that current density at the center part of the conductors reaches near to zero but at the skin parts there are very much current density values. Fig. 8 (b) and (d) show the skin effect in a cross section of one conductor, explicitly. In these two figures, at both circular and square shape conductors, the current density decreases with a steep slope to almost zero and again increase with a fast slope to around 70 A/mm$^2$ and 30 A/mm$^2$ for circular and square shape conductors, respectively. It seems that current density in the AWG18 circular shape is much higher than the current density at the square shape conductor that is equivalent to the same AWG.

Also, the numeric variations of the current density for both types of the transformers at 20 kHz are displayed in Fig. 9 with the same condition for the transformer as 20 MHz, to compare the influence of the skin effect on both types.

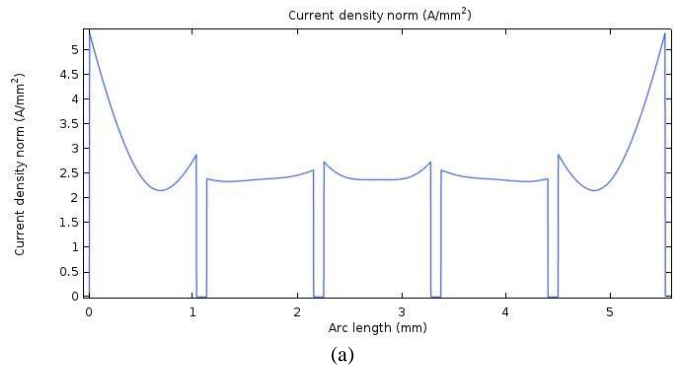

(a)

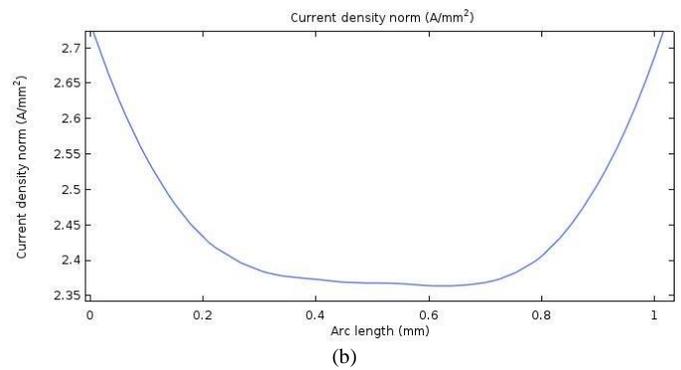

(b)

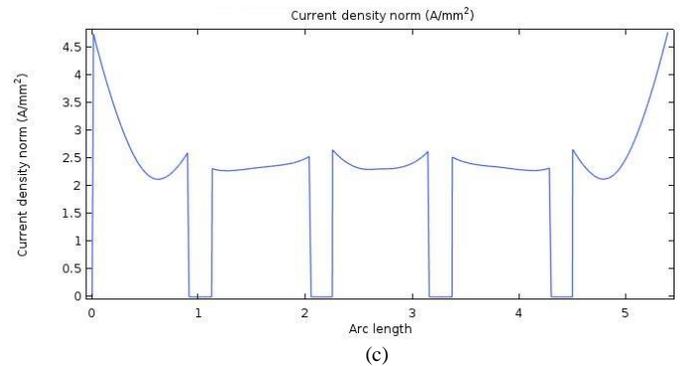

(c)

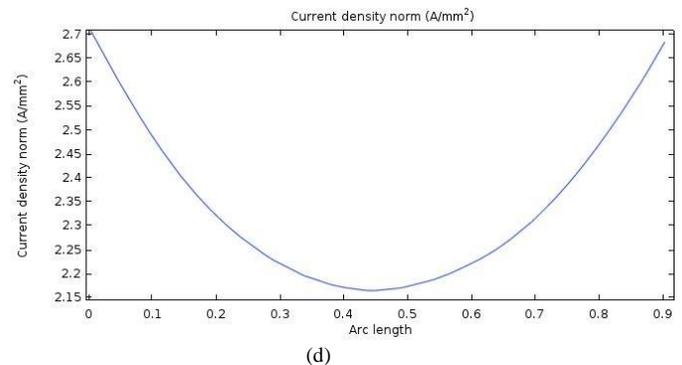

(d)

Fig. 9. Current density norm (A/mm$^2$) across the defined cutline in Fig. 7 from bottom to above at 20 kHz. a) Current density on the cutline of Fig.7 (a) across the primary conductors at the right window. b) Current density on the cutline of Fig.7 (b) across the primary conductors at the right window. c) Current density on the cutline of Fig.7 (c) across the primary conductors at the right window. d) Current density on the cutline of Fig.7 (d) across the primary conductors at the right window.

It is obvious from the Fig. 9 that current density distribution is more even, in comparison to the 20 MHz and there is remarkable current density at the center parts of the conductors, which is expected to lead more leakage inductance and less windings AC resistance.

The maximum of current density in the conductors at 20 kHz is much lower than maximum of the current density at 20 MHz and the distribution is very even. Also, it is clear that in the square shape conductors, the skin effect is slightly weaker than the AWG18 circular shape conductors. This feature can help in the transformers with high number of turn windings to reduce the skin effect, even marginally; because it may lead to less winding losses. Furthermore, in some applications, which may use litz wire to reduce skin effect, choosing the shape of the strands can lead to lowering litz wire loss.

## B. Leakage Inductance Calculations

To find the leakage inductance for all types of the transformers, by considering (5) it is clear that magnetic field distribution strength (H) is a significant factor. H distribution as the key factor of calculating leakage inductance by short circuiting the secondary for all four types of the transformer are depicted in Fig. 10. Also, magnetizing inductance, leakage inductance, and AC resistance of the windings seen on the primary side, are displayed in TABLE III by considering the skin effect at 20 MHz; it is noted that, as the magnetic field strength get intensified in the space-especially in the window area-the leakage inductance goes higher.

To compare the influence of the skin effect on the results, the leakage inductance and AC resistance at 20 KHz are shown in TABLE IV. As was expected, the leakage inductance changes directly and the AC resistance changes indirectly with the frequency due to the skin effect and eddy current. In all of the transformer types this pattern is observable, except type c. Because the structure and the winding arrangement of type c, cancel the skin effect and prevent the eddy current to flow.

TABLE III shows a range of different values for the leakage inductance and the AC resistance in each of the transformer types. Lowering the parasitic parameters and loss parameters help to a better performance for the power electronics devices, in very high frequency. Also, changing the parameters can lead to remove unwanted resonant frequencies. As it is clear in this table, by using the foil conductors, AC resistance reduces, up to 400 times for the type c (overlaid foil windings) and 12 times for the type d (non-overlaid foil windings), compared to the AWG conductors. The leakage inductance in the AWG circular shape conductors is 2 times greater than the square shapes. By using the foil conductors of the type c, in spite of lowering winding loss, the leakage inductance is 2 times and 4 times greater than the AWG circular shape and the square shape conductors, respectively. Also, using foil conductors of type d decreases winding loss significantly too, but in this case the leakage inductance is 8 times, 14 times, and 4 times greater than the AWG circular shape, square shape conductors and foil of the type c, respectively. It is noted that the type d AC resistance is 30 times greater than the type c. Hence, deciding to choose which topology and structure needs a compromise between parameters.

Fig. 11 illustrates the leakage inductance and the winding AC resistance of the transformers at the different frequencies from 20 kHz to 20 MHZ. Fig 11 (c) shows the absence of the eddy current in the transformer windings at all frequencies. Fig 11 (d) shows that at the frequencies higher than 200 kHz eddy currents cannot flow in the foil windings of this transformer, while at the lower frequencies, the windings show some portions of eddy currents.

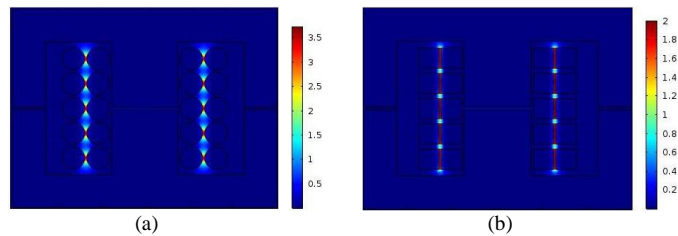

(a)       (b)

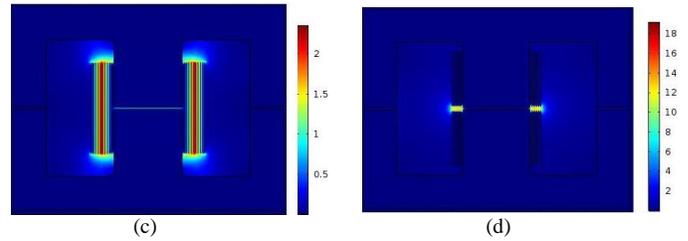

(c)       (d)

Fig. 10. Magnetic field strength (A/mm) at 20 MHz. a) Windings with the AWG18 circular shape conductors b) Windings with the square shape conductors with the same cross sectional area to the AWG18. c) Type (I) of windings with foil conductors. d) Type (II) of windings with foil conductors.

TABLE III
Magnetizing inductance, leakage inductances, and windings AC resistance of the transformers at 20 MHz. a,b,c,d types are the same as Fig. 4.

| Transformer Type | Magnetizing Inductance(nH) | Leakage Inductance(nH) | AC Resistance(mΩ) |
|---|---|---|---|
| (a) | 397.60 | 6.8558 | 80.637 |
| (b) | 397.08 | 3.8030 | 73.377 |
| (c) | 413.33 | 13.2631 | 0.19381 |
| (d) | 422.14 | 51.765 | 6.7345 |

TABLE IV
Leakage inductances and windings AC resistance of the transformers at 20 kHz. a,b,c,d types are the same as Fig. 4.

| Transformer Type | Leakage Inductance(nH) | AC Resistance(mΩ) |
|---|---|---|
| (a) | 22.910 | 1.8264 |
| (b) | 19.021 | 1.8521 |
| (c) | 13.451 | 0.10297 |
| (d) | 63.390 | 1.6339 |

Also, in spite of observing almost the same magnetizing inductance in all of the four types, in all frequencies, these graphs show, the nonlinear behavior of the skin effect can lead to different parasitic and winding loss behavior.

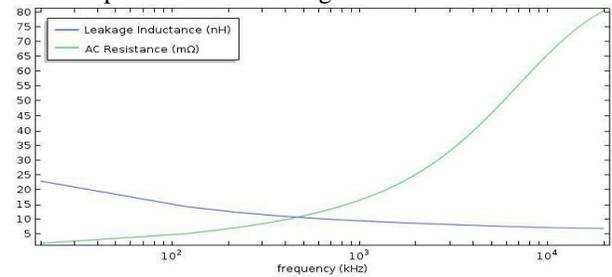

(a)

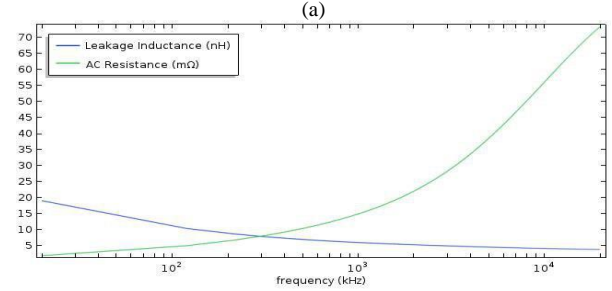

(b)

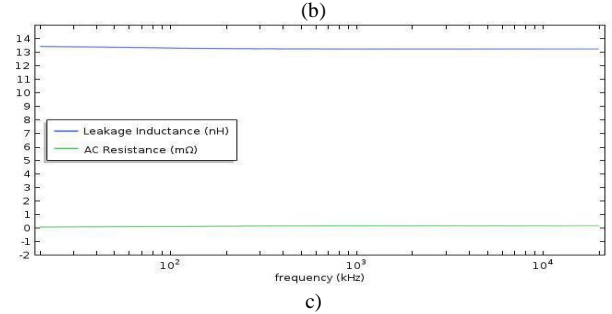

c)

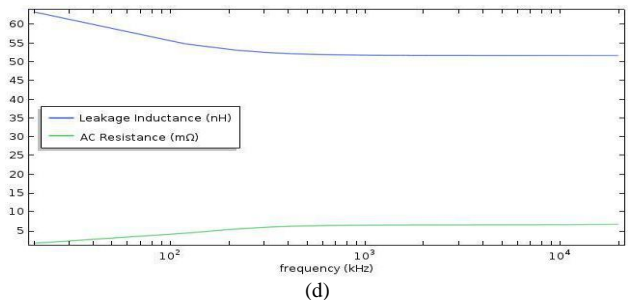
(d)

Fig. 11. Leakage inductance and winding AC resistance at different frequencies. a) Windings with the AWG18 circular shape conductors b) Windings with the square shape conductors with the same cross sectional area to the AWG18. c) Type (I) of the windings with the foil conductors. d) Type (II) of the windings with the foil conductors.

## IV. CONCLUSION

The skin effect is responsible for leakage inductance decrement and winding AC resistance increment at higher frequencies. In conductors such as circular and square shaped wires, which are studied in this paper, the skin effect leads to variations in the leakage inductance and the winding losses. To mitigate this, foil conductors are a suitable choice for the conductor shape. By using foil conductors, the winding arrangements in the transformers play an effective role as well. As it is shown in the previous section, the winding arrangements, the conductor shapes, and structures can change the results of parasitic and loss behavior of the high frequency transformers significantly, while giving the same value for the magnetizing inductance. According to the results of this study for the same value of the magnetizing inductance, the transformer structure and the winding topology can change the leakage inductance and the winding AC resistance up to 13 times and 400 times, respectively. Moreover, in the wire conductors, shape of the wires despite of the same value for cross sectional area, can affect the parasitic and loss behavior. The reason is that, this shape can slightly change the skin effect behavior in the conductors. In this study, by changing the square shape to the circular shape the leakage inductance and the winding loss increase up to 2 times and 1.2 times, respectively. Furthermore, the frequency responses for all of the winding structures studied, show that in wire conductors, the leakage inductance and the winding AC resistance change significantly between lower and higher frequencies. However, this behavior is not noticeable in the foil conductors. Lastly, to choose the best structure, winding arrangement, and conductor at very high frequency domain, for having a better parasitic and loss behavior, it is essential to make a trade-off in designs.

## ACKNOWLEDGEMENT

This material is based upon work supported by the U.S. Department of Energy, "Enabling Extreme Fast Charging with Energy Storage", DE-EE0008449.

## REFERENCES

[1] Perreault, David J., et al. "Opportunities and Challenges in Very High Frequency Power Conversion." *2009 Twenty-Fourth Annual IEEE Applied Power Electronics Conference and Exposition*, 2009, doi:10.1109/apec.2009.4802625.
[2] Shamsi, Pourya, and Babak Fahimi. "Design and Development of Very High Frequency Resonant DC–DC Boost Converters." *IEEE Transactions on Power Electronics*, vol. 27, no. 8, 2012, pp. 3725–3733., doi:10.1109/tpel.2012.2185518.
[3] S. Yazdani and M. Ferdowsi, "Robust Backstepping Control of Synchronverters under Unbalanced Grid Condition," *2019 IEEE Power and Energy Conference at Illinois (PECI)*, 2019.
[4] S. Yazdani and M. Ferdowsi, "Voltage Sensorless Control of a Three-Phase Standalone Inverter Based on Internal Model Control," *2018 IEEE Energy Conversion Congress and Exposition (ECCE)*, 2018.
[5] S. Yazdani and M. Ferdowsi, "Hardware Co-simulation of Voltage Sensorless Current Control Based on Internal Model Principle," *2018 IEEE Energy Conversion Congress and Exposition (ECCE)*, 2018.
[6] J.-M. Choi, B.-J. Byen, Y.-J. Lee, D.-H. Han, H.-S. Kho, and G.-H. Choe, "Design of Leakage Inductance in Resonant DC-DC Converter for Electric Vehicle Charger," *IEEE Transactions on Magnetics*, vol. 48, no. 11, pp. 4417–4420, 2012.
[7] C. W. T. McLyman, *Transformer and inductor design handbook*. CRC Press, 2017.
[8] Fair-rite products crop, 67 material datasheet, Costa Mesa, CA, USA [online]. Available: https://www.fair-rite.com/67-material-data-sheet/